\documentclass[conference]{IEEEtran}
\IEEEoverridecommandlockouts
\usepackage{cite}
\usepackage{amsmath,amssymb,amsfonts}
\usepackage{algorithmic}
\usepackage{graphicx}
\usepackage{textcomp}
\usepackage{xcolor}
\usepackage{multirow}
\usepackage{listings}
\usepackage{parcolumns}
\usepackage{url}
\def\BibTeX{{\rm B\kern-.05em{\sc i\kern-.025em b}\kern-.08em
    T\kern-.1667em\lower.7ex\hbox{E}\kern-.125emX}}

\definecolor{codegreen}{rgb}{0,0.6,0}
\definecolor{codegray}{rgb}{0.5,0.5,0.5}
\definecolor{codepurple}{rgb}{0.58,0,0.82}
\definecolor{backcolour}{rgb}{0.95,0.95,0.92}

\lstdefinestyle{mystyle}{
    backgroundcolor=\color{backcolour},   
    commentstyle=\color{codegreen},
    keywordstyle=\color{magenta},
    numberstyle=\tiny\color{codegray},
    stringstyle=\color{codepurple},
    basicstyle=\ttfamily\footnotesize,
    breakatwhitespace=false,         
    breaklines=true,                 
    captionpos=b,                    
    keepspaces=true,                 
    numbers=left,                    
    numbersep=5pt,                  
    showspaces=false,                
    showstringspaces=false,
    showtabs=false,                  
    tabsize=2
}
\begin{document}

\title{Portability for GPU-accelerated molecular docking applications for cloud and HPC: can portable compiler directives provide performance across all platforms?
\thanks{This manuscript has been authored by UT-Battelle, LLC under Contract No. DE-AC05-00OR22725 with the U.S. Department of Energy. The United States Government retains and the publisher, by accepting the article for publication, acknowledges that  the  United  States  Government  retains a non-exclusive, paid-up, irrevocable, world-wide license to publish or reproduce the published form of this manuscript, or allow others to do so, for United States Government purposes. The Department of Energy will provide public access to these results of federally sponsored research in accordance with the DOE Public Access Plan (http://energy.gov/ downloads/doe-public-access-plan).}
}

\author{
    \IEEEauthorblockN{
    Mathialakan Thavappiragasam\IEEEauthorrefmark{1}, 
    Wael Elwasif\IEEEauthorrefmark{1}, 
    Ada Sedova\IEEEauthorrefmark{1},
    \IEEEauthorblockA{\IEEEauthorrefmark{1}Oak Ridge National Laboratory, Oak Ridge, TN
    \\
    Corresponding email: {[thavappiragm, elwasifwr, sedovaaa]@ornl.gov}
    }
    }
    }

\maketitle
\thispagestyle{plain}
\pagestyle{plain}
\begin{abstract}
High-throughput structure-based screening of drug-like molecules has become a common tool in biomedical research. Recently, acceleration with graphics processing units (GPUs) has provided a large performance boost for molecular docking programs. Both cloud and high-performance computing (HPC) resources have been used for large screens with molecular docking programs; while NVIDIA GPUs have dominated cloud and HPC resources, new vendors such as AMD and Intel are now entering the field, creating the problem of software portability across different GPUs. Ideally, software productivity could be maximized with portable programming models that are able to maintain high performance across architectures. While in many cases compiler directives have been used as an easy way to offload parallel regions of a CPU-based program to a GPU accelerator, they may also be an attractive programming model for providing portability across different GPU vendors, in which case the porting process may proceed in the reverse direction: from low-level, architecture-specific code to higher-level directive-based abstractions. MiniMDock is a new mini-application (miniapp) designed to capture the essential computational kernels found in molecular docking calculations, such as are used in pharmaceutical drug discovery efforts, in order to test different solutions for porting across GPU architectures. Here we extend MiniMDock to GPU offloading with OpenMP directives, and compare to performance of kernels using CUDA, and HIP on both NVIDIA and AMD GPUs, as well as across different compilers, exploring performance bottlenecks. We document this reverse-porting process, from highly optimized device code to a higher-level version using directives, compare code structure, and describe barriers that were overcome in this effort.
\end{abstract}

Drug discovery is a lengthy process; computational efforts aid bench-top scientists search for possible small molecule therapeutics to test experimentally \cite{drews2000drug,pagadala2017software}. Computational approaches are much faster and cheaper than experimental methods for filtering combinations of compounds and proteins, and can help to decrease the size of the chemical space that must be searched for lead compounds. Chemical synthesis companies today promise the ability to synthesize over a billion different molecules that could serve as lead compounds for optimization as drugs. By acting as a screening tool, computational molecular docking, an approach that simulates  the three dimensional interactions of small molecules with a target protein \cite{pagadala2017software,vermaas2020supercomputing,legrand2020gpu}, can reduce this space to a reasonable subset for experimental testing. Recently, high-throughput molecular docking has made use of large, parallel computing resources to perform extremely large screens, from millions to billions of small molecular ligands against proteins \cite{gorgulla2020open,glaser2021high}. 

\begin{figure}[htb]
\centering{\includegraphics[ width=0.8\linewidth]{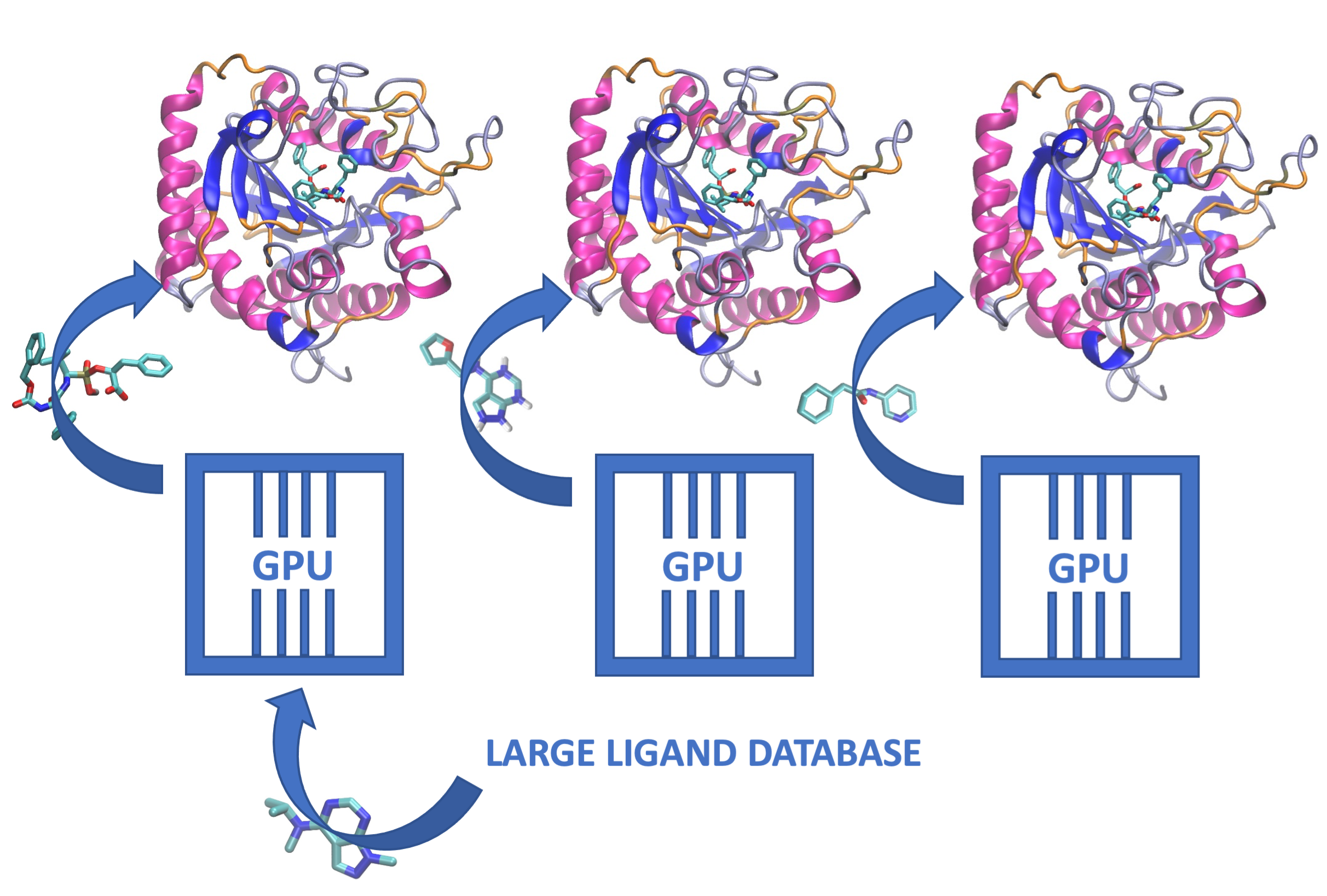}}
\caption{Scheme of the high-throughput GPU-accelerated molecular docking workflow. A large set of small molecule drug-like ligands (tens of thousands to billions) is rapidly screened using all available GPUs on a cloud or cluster resource to find an optimal position within a target protein active site.}
\label{fig:flow}
\vspace{-1em}
\end{figure}
As part of this work, acceleration with graphics processing units (GPUs) has been recently shown to provide a dramatic speedup for molecular docking programs \cite{legrand2020gpu,glaser2021high,santos2019accelerating}. Figure~\ref{fig:flow} illustrates the workflow for high-throughput structure-based screens of small molecule ligand databases with distributed GPU-accelerated molecular docking; NVIDIA provides the AutoDock-GPU program in its NGC Catalog of containeerized software for servers.\footnote{\url{https://catalog.ngc.nvidia.com/orgs/hpc/containers/autodock}} The high-throughput parallel deployment of docking on large set of heterogeneously-sized input ligands also presents a load balancing and dataflow execution optimization problem \cite{thavappir2021addressing}.

Computer architectures are changing rapidly. High-performance computing (HPC) has increasingly made use of heterogeneous multi-node systems with accelerators such as GPUs, and parallelization over multicore central processing units (CPUs), to provide optimal performance for applications in both leadership computing facilities and on cloud and data-center resources. Likewise, distributed cloud resources have also made extensive use of GPUs, especially for deep learning applications. Over the past decade, much effort has gone into programming GPUs using device-specific APIs and optimizations to maximize performance. Recent emergence of several GPU vendors competing for cloud and HPC users has upended this programming paradigm: optimizing and maintaining multiple versions of a program for each GPU vendor quickly becomes intractable. Therefore, portable programming solutions that can provide performance as close to optimized architecture-specific versions as possible are becoming more urgent needs for maintaining productivity in HPC \cite{pennycook2019implications, harrell2018effective,sedova2018using,sedova2018high, thavappiragasam2020performance}. 
 
 Mini-applications (miniapps) are an essential tool used by the HPC community to test and optimize both performance and portability of important HPC-based programs. Miniapps aim to capture the key kernels used in larger applications to test program performance and portability, and to aid with transitions to these new systems \cite{messer2018miniapps,thavappiragasam2020performance}.
  We have developed a miniapp which focuses on the computational kernels used in molecular docking applications. 
 MiniMDock\footnote{https://github.com/ORNL-PE/miniMDock}$^,$\footnote{https://proxyapps.exascaleproject.org/app/minimdock/} is based on an application that was adapted for high-throughput computational screening of potential drug leads on the Summit supercomputer \cite{legrand2020gpu} and was used to dock over a billion molecules from a drug-discovery-compound dataset to a viral protein for pharmaceutical therapeutics development for COVID-19 \cite{glaser2021high}.
 
 OpenMP is a portable directive-based parallel programming API used ubiquitously in computational applications to program CPUs. Recently, compiler-directive-based parallelization offloaded to GPU devices has become possible with OpenMP target offload~\cite{Openmp5} using OpenMP 4.5/5.1 compilers. These directives would, in principle, be completely portable to any GPU regardless of vendor as long as a compiler backend exists targeting that device. The main challenges are then 1) for compiler developers to provide high-performing implementations capable of optimizing many types of code patterns and 2) for application developers to develop an optimized code structure that maximizes performance over multiple GPU devices. Here, we explore both of these challenges as we extend the miniMDock miniapp to OpenMP target offload, testing performance using both NVIDIA and AMD GPUs with several compilers per architecture, and compare to the performance of device-specific code which for both types of devices, makes use of low-level optimizations.
 
\section{Background}
MiniMDock provides a testing application with an algorithmic pattern different from many other HPC miniapps \cite{messer2018miniapps}. The calculation, which is ultimately an optimization over a potential-energy surface described by a particle-grid interaction, is based on a genetic algorithm. While MiniMDock's kernels are adapted from our work \cite{legrand2020gpu} porting the relatively new, GPU-accelerated AutoDock-GPU program \cite{santos2019d3r,santos2019accelerating} from OpenCL to CUDA and applying modifications for high-throughput use in screening large numbers of small-molecule compounds (ligands) against a protein receptor, the algorithm and the original implementation on CPUs was developed over several decades. 

Multiple strategies and solutions exist today aimed at achieving performance portability between varying GPU architectures and across a wider set of parallel programming targets such as CPU-based threading. OpenCL was designed to provide parallel programming solutions for both CPU threads (and SIMD) and GPU accelerators \cite{stone2010opencl}, and is supported on NVIDIA GPUs, AMD GPUs, and FPGAs.\footnote{\url{https://www.khronos.org/registry/OpenCL/}} The Kokkos middleware API\footnote{\url{https://github.com/kokkos/kokkos}} is facilitated by C++ template libraries and has been focused on portability across many parallel architectures as well, including CPU threads and GPUs, with support for AMD GPUs currently being provided through an interface with the HIP API.\footnote{\url{https://rocmdocs.amd.com/en/latest/ROCm_API_References/HIP-API.html}} Beginning close to a decade ago, options for directive-based GPU offloading have appeared, supported by several different compilers, and using two main solutions: OpenACC and OpenMP.

In the following subsections, we provide further details on the molecular docking algorithm used here, the structure of the miniapp, and the use of directives for GPU programming.

\subsection{Design and Structure of MiniMDock}

\begin{figure}[htbp]
\centerline{\includegraphics[ width=0.5\linewidth]{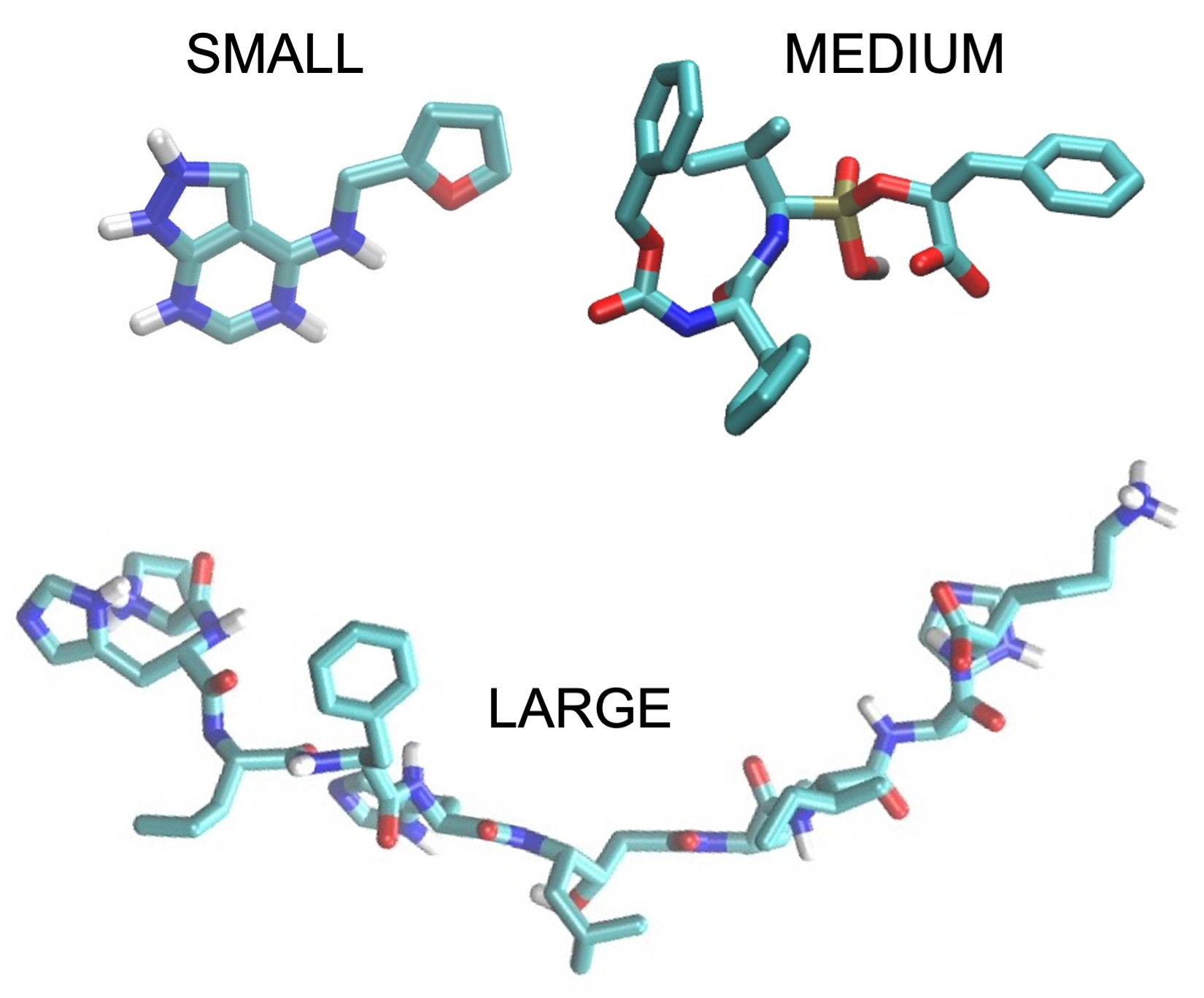}}
\caption{The three input sizes for the MiniMDock miniapp.}
\label{fig:ligs}
\vspace{-1em}
\end{figure}

Briefly, the program represents the protein-ligand interaction with a reaction field for the protein portion, and with an particle representation for the ligand. The reaction field is represented by a three-dimensional grid. The ligand atoms are represented by points in space that, along with the effects of the the interaction grid, are subject to self-interactions such as hydrogen bonding, van der Waals forces, and desolvation effects. The algorithm is an energy minimization problem designed to find the most energetically favorable pose for the ligand within the protein's binding pocket. The genetic algorithm uses a set of ``crossover" events  where features of each solution are mixed and interchanged across a ``population" of possibilities, and individuals in the population are selected for the next iteration based on a numerical fitness score. The algorithm also incorporates a memetic component, wherein a local optimization is performed on a random sample of individuals of each generation, with the possibility to pass on fitness benefits obtained during the local search to the offspring during the crossover. The local optimization method is based on a random optimizer called the Solis-Wets algorithm \cite{solis1981minimization}. The docking algorithm parallelizes well on GPUs, as each local optimization occurs independently from other members of the population. As an additional level of parallelism, several full optimizations each with a starting population of 150 individuals are computed simultaneously. The final solution is the best scoring pose out of all final solutions over all runs. 

For the miniapp, several file writing tasks were removed from the program, to focus on the time spent in compute kernels. Three different inputs are provided for testing: a small, medium, and large task. These differ by the number of atoms in the ligand, which are used in the calculation of the interaction energy, the number of rotatable bonds, which together with translations of the ligand are the degrees of freedom for the optimization \cite{thavappiragasam2020performance}. The small, medium, and large inputs have 21, 43, and 108 atoms, and 2, 15, and 31 rotatable bonds, respectively, and are shown in Figure~\ref{fig:ligs}. 

\begin{figure}[htbp]
\centerline{\includegraphics[ width=0.9\linewidth]{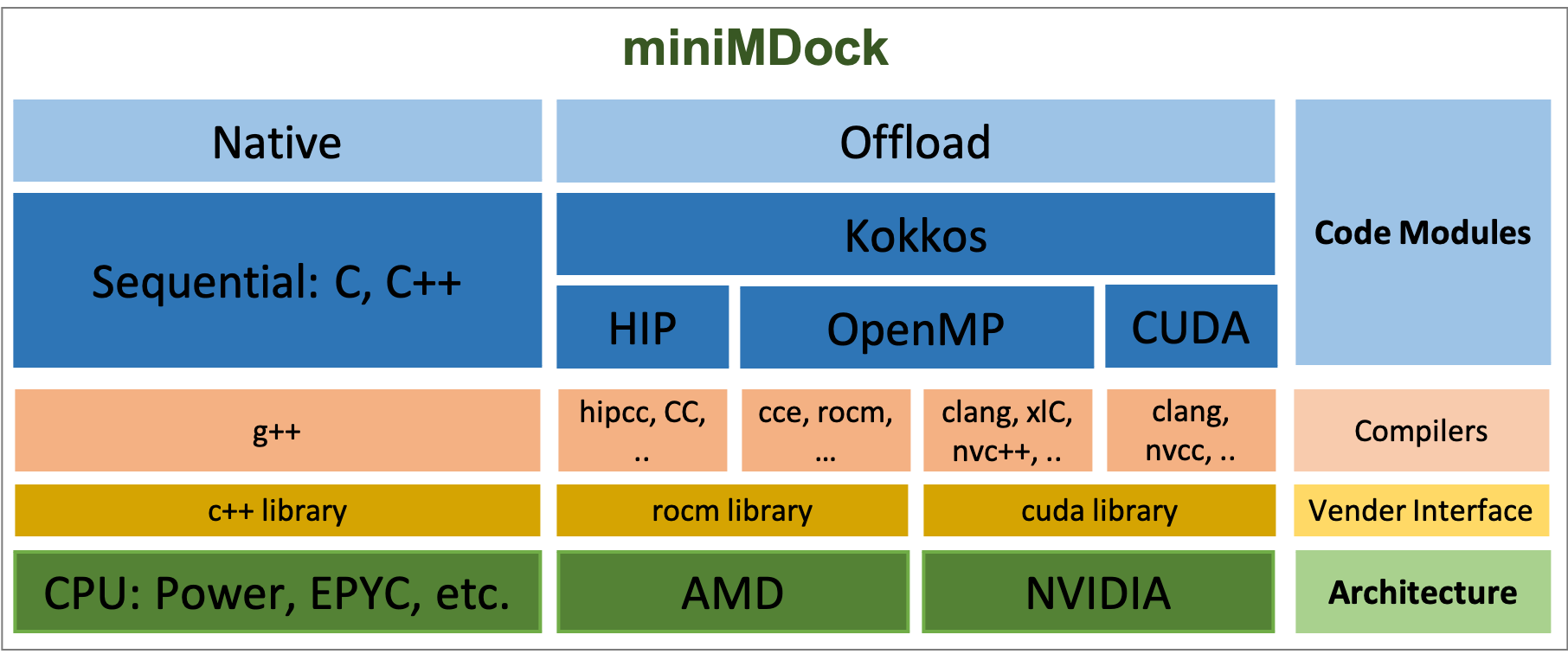}}
\caption{Software stack of the miniapp MiniMDock.}
\label{fig:sw_stack}
\vspace{-1em}
\end{figure}
Figure~\ref{fig:sw_stack} shows the software-stack layers in the MiniMDock miniapp. Here we focus on comparing the new OpenMP target module to the two versions that use the device specific APIs: CUDA for NVIDIA GPUs, and HIP for AMD GPUs. While HIP is designed to be portable to CUDA devices, it has been developed by AMD to provide optimized performance on AMD GPUs, so here we treat it as the most hardware-specific programming model for AMD GPUs.

\subsection{Directives for GPU Offloading}
 OpenMP began as a CPU-threading programming model and began offering offloading to GPUs in version 4.0; it has continued to develop support for new features to date \cite{Openmp5}. OpenACC was originally supported only by the PGI compiler, which was recently acquired by NVIDIA and now is distributed as part of NVIDIA HPC SDK (NVHPC). OpenMP offloading for NVIDIA GPUs is currently supported by the IBM XL compiler, LLVM, and NVHPC, while support for AMD GPUs has recently been provided by HPE Cray's Programming environment, specifically the Cray Compiling Environment (CCE), and by AMD's ROCm environment, in preparation for the Frontier supercomputer soon to be deployed at the Oak Ridge Leadership Computing Facility (OLCF).

While OpenACC was created for a ``descriptive" programming style, wherein the compiler is responsible for most decisions about the way a parallel region is offloaded and distributed across the GPU, OpenMP was traditionally designed for a ``prescriptive" model, where the application developer provides specific instructions for mapping parallel regions to the architecture. The prescriptive model thus can create code that may not be as performance portable across architectures. However, research efforts have focused on understanding how OpenMP can be used in a less architecture specific manner to improve portability \cite{juckeland2016describing}. Performance gaps between optimized CUDA code and directive-based versions have been narrowing in recent years, for both OpenACC and OpenMP \cite{sedova2018high,boehm2018evaluating,daley2020case}, paving the way for more large applications to invest time in supporting a directive-based version or to chose a directive-based offloading strategy as the primary programming model for GPU support, such as was recently done for the widely-used materials program VASP,\footnote{\url{https://www.vasp.at/wiki/index.php/OpenACC_GPU_port_of_VASP}} and the polarizable molecular dynamics program Tinker-HP \cite{adjoua2021tinker}.

\section{Related Work}
Numerous studies have focused on directive-based offloading as a solution for performance portability over the past decade \cite{juckeland2016describing,gayatri2018case,daley2020case,lopez2016towards,martineau2016pragmatic, sedova2018using,hsu2018performance}. Multiple reports compared the OpenACC and OpenMP approaches and explored difference in usage and performance on GPUs compared to CUDA, and for CPU-based threading and accelerators like the Xeon Phi \cite{juckeland2016describing,gayatri2018case,lopez2016towards,boehm2018evaluating}, using simplified kernels and miniapps. Our miniapp contains examples of compute patterns different from other commonly used miniapps for HPC. Here we present one of the first studies, to our knowledge, of performance portability for directives across NVIDIA and AMD GPUs, using several compilers, with a new module of this miniapp. We also describe the reverse-porting of a highly optimized architecture-specific code to a directive-based code, and explore how this endeavor affects the final program and the performance, while providing guidance based on our experiences.

\section{Porting From Optimized CUDA}

The starting point for the port to OpenMP target offload is a derivative of a hand-optimized version resulting from a collaboration between NVIDIA, Scripps Research and the Oak Ridge National Lab. The final product included warp-level primitives and the ability to utilize different numbers of threads per block to tune performance \cite{legrand2020gpu,thavappiragasam2020performance}. Code restructuring was therefore required for the OpenMP target version. This resulted in what may be a different final product from what may have come from either a \textit{de novo} programming effort or from applying OpenMP offloading to a serial CPU version or one with CPU threading. One of the main differences between programming using OpenMP directives versus using architecture-specific code is that directives provide a more high-level set of instructions to the compiler; the compiler and the runtime have more flexibility as to how to perform the actual operation. Directives such as OpenMP allow the developer to express code and the control flow in the problem domain, as opposed to the explicit mapping of operations to specific threads; the decisions about how to assign computations to the hardware are made by the compiler and runtime. Lower-level APIs such as CUDA and HIP require the developer to express the problem in terms of the hardware. One of the key challenges in porting from architecture-specific code is to thoroughly understand the difference between the programming models, and what can be expected in both: where the compiler has flexibility, and how parallel constructs such as synchronization and barriers must be implemented. This understanding can be crucial for correct program behavior. The lower-level code already lays out the parallelism, and to reduce re-writing, some decisions regarding design of these regions may potentially produce sub-optimal performance for the directive-based version. Another key difference between low-level massively-parallel programming models such as CUDA and HIP and the OpenMP programming model is the availability of a hierarchy of parallel constructs in OpenMP that is not available natively in CUDA. In CUDA, a kernel is executed by a single thread, and higher level constructs (e.g. block-level operations) need to be implemented in the kernel code itself, with no support directly from the CUDA programming model. The OpenMP {\tt teams} construct maps naturally to block-level operations (as {\tt teams} parallelism is implemented using a single thread per CUDA block or HIP workgroup). This difference manifests itself in the need to restructure a complex kernel where certain code regions are done by a single thread. In OpenMP, a sequence of parallel regions where all threads in the team are participating in a (typically) worksharing construct, using {\tt threadId}, interleaved with code regions where team-level parallelism is in effect (and a single thread in the team is active), can help to facilitate this translation. In this section we demonstrate how the architecture-specific CUDA/HIP code was restructured to address these differences. 

\subsection{Removing architecture-specific code}

For OpenMP code to be highly portable, it needs to avoid the use of architecture-specific APIs and low-level primitives that are not universally available. In CUDA and HIP, low-level primitives are used to perform operations that typically involve participation by threads in an entire block (these are typically referred to as warp-level primitives).
Listings~\ref{code:sumevals_cuda} and~\ref{code:sumevals_ompt} show an example of the restructuring of a parallel region going from a mapping in CUDA or HIP to threads and blocks, followed by a warp-level reduction, to the use of OpenMP teams within a target region, and an inner {\tt parallel for} reduction. In both versions, the structure {\tt cData} is accessed to obtain the required information for each member of a population in the genetic algorithm, across the {\tt nruns} independent replicas. In both versions, a flattened array is also used, although this was done to reduce code restructuring (and thus increase productivity), and not because this is required by OpenMP. 

\lstset{language=C++,basicstyle=\footnotesize,numberstyle=\tiny,frame=single,caption={The sum\_evals kernel in CUDA; HIP version is very similar.},label=code:sumevals_cuda} %,xleftmargin=0.3cm , linewidth=8.8cm}
\begin{lstlisting}
__global__ void gpu_sum_evals_kernel(...){
    __shared__ int sSum_evals;
	int partsum_evals = 0;
    int* pEvals_of_new_entities = cData.pMem_evals_of_new_entities + blockIdx.x * cData.dockpars.pop_size;
  	for (int entity_counter = threadIdx.x;
            entity_counter < cData.dockpars.pop_size;
            entity_counter += blockDim.x){
            partsum_evals += pEvals_of_new_entities[entity_counter];
	}
    // warp level reduction
    REDUCEINTEGERSUM(partsum_evals, &sSum_evals);
    // master thread assigns final value
    if (threadIdx.x == 0){
        cData.pMem_gpu_evals_of_runs[blockIdx.x] += sSum_evals;
    }
} // end for the kernel
\end{lstlisting}

%\lstset{language=C++,basicstyle=\footnotesize,numberstyle=\tiny,frame=single,caption={Macro for CUDA warp-level reduction},label=code:Reduc_Fun_CUDA}
%\lstinputlisting{listings/warp_reduc.cpp}

%\lstset{language=C++,basicstyle=\footnotesize,numberstyle=\tiny,frame=single,caption={Macro for HIP wavefront-level reduction },label=code:Reduc_Fun_HIP}
%\lstinputlisting{listings/wavefront_reduc.cpp}
\lstset{language=C++,basicstyle=\footnotesize,numberstyle=\tiny,frame=single,caption={The sum\_evals kernel in OpenMP target offloading.},label=code:sumevals_ompt}
\begin{lstlisting}
void gpu_sum_evals(uint32_t nruns, uint32_t work_pteam, ... ){
#pragma omp target teams distribute\
    num_teams(nruns) thread_limit(work_pteam)
    for (int idx = 0; idx < nruns; idx++ ) {
        int sum_evals = 0;
        int* pEvals_of_new_entities = cData.pMem_evals_of_new_entities + idx * dockpars.pop_size;
    // team level reduction
#pragma omp parallel for reduction(+:sum_evals)
        for (int j = 0; j < work_pteam; j++){
            int partsum_evals = 0;
            for (int entity_counter = j; 
                entity_counter < dockpars.pop_size;
                entity_counter +=work_pteam){
                partsum_evals += pEvals_of_new_entities[entity_counter];
	        }
            sum_evals += partsum_evals;
        }// end for a team
        // master thread assigns final vale
        cData.pMem_gpu_evals_of_runs[idx] += sum_evals;
   }// end for a set of teams
} // end for the kernel
\end{lstlisting}

 The HIP and CUDA versions demonstrate how only one level of parallelism is provided by these APIs. The reduction in these kernels use warp-level primitives \texttt{\_\_shfl\_sync} and \texttt{\_\_any\_sync} \cite{thavappiragasam2020performance}. For the same reduction, OpenMP uses a team-level reduction as shown between lines 9 and 18 in Listing~\ref{code:sumevals_ompt}. The individual replicas ({\tt nruns}) are worked on at the teams level in an execution pattern controlled by the runtime. This ``team-level" code is implemented in CUDA with code executed by {\tt threadIdx=0} corresponding to team level parallelism in OpenMP, it is an explicit construct provided by the programming model. OpenMP provides support for high-level operations that involve threads in a parallel region through the {\tt reduction} construct with many pre-defined reduction operators that are part of the OpenMP standard. The combined use of OpenMP reduction and OpenMP team-level parallel regions allows the desired operation to be expressed at a higher level, while leaving it up to the OpenMP compiler implementation to generate the most efficient {\em inter-thread} code that may involve the use of low level primitives for the target platform.

We implemented two different programmatic approaches for the OpenMP target offload and evaluated their performance portability features; in general, a good compiler-directive implementation should be relatively insensitive to variations in program structure and control flow, but in practice, this may not be the case. Therefore, our tests for performance portability also include this aspect.

\subsection{Approach 1: Using ``teams distribute" construct}

In this approach (which was introduced above), we use the higher-level OpenMP constructs \texttt{distribute} and \texttt{parallel for} to automatically distribute the work over threads, without directly specifying thread-level work assignments. The outer-loop is distributed over teams using \texttt{distribute} construct and the inner-loop is shared over the threads using \texttt{parallel for} construct. However, extra effort is required for dealing with thread synchronization in this approach. 

\lstset{language=C++,basicstyle=\footnotesize,numberstyle=\tiny,frame=single,caption={Thread synchronization in CUDA.},label=code:synchthread_cuda}
\begin{lstlisting}
__global__ void gpu_kernel( parameters ... ){
	__shared__ float3 A[N];
	float  x = device_function( A, ... );
	__syncthreads();
	compute( ... );
	if (threadIdx.x == 0) { 
	    //Work for the master thread
	}
}
void use_gpu_kernel(uint32_t nblocks, uint32_t threads_pblock, parameters ...){
    gpu_kernel<<<nblocks, threads_pblock>>>( parameters ...);
}
\end{lstlisting}

\lstset{language=C++,basicstyle=\footnotesize,numberstyle=\tiny,frame=single,caption={Thread synchronization in OpenMP target offloading using \textit{Approach 1}.},label=code:synchthread_ompt}
\begin{lstlisting}
void gpu_kernel( uint32_t nteams, uint32_t threads_pteam, parameters ... ){
#pragma omp target teams distribute num_teams(nteams) thread_limit(threads_pteam)
	for (int i = 0; i < nit_atteam; i++){
	    float3_struct A[N];	
#pragma omp parallel for
	    for (int j = 0; j < work_pteam; j++){
	        float  x = device_function( A, ... );
	    }// end of a team -- implicit barrier
#pragma omp parallel for
	    for (int j = 0; j < work_pteam; j++){
	        compute( ... );
	    }// end of a team
	    { ... } //Work for the master thread
	}// end of teams
}
\end{lstlisting}

Thread synchronization ensures that no race conditions occur during write/read accesses and make the current (correct) values visible for each thread over the period of execution. Different programming models use different synchronization functions to synchronize runtime activities based on the architecture. CUDA uses the function \texttt{\_\_synchthreads()} to place thread-level barriers and coordinate communication among threads in the same block. Similarly, memory-fence functions enforce the ordering of memory accesses, depending on the scope in which the ordering is enforced. The CUDA memory-fence function \texttt{\_\_threadfence()} works beyond the thread block and ensures that there are no race condition for any threads on the device. OpenMP provides an {\tt omp barrier} directive for thread synchronization, but unfortunately this directive cannot be used to synchronize across teams. However, an implicit barrier exists for a team construct when a \texttt{omp parallel for} directive is nested within the \texttt{omp target team distribute} directive.

Listings~\ref{code:synchthread_cuda}  and~\ref{code:synchthread_ompt} show the differences between CUDA and OpenMP code for a region that requires a barrier and thread synchronization (line 4 in Listing~\ref{code:synchthread_cuda}). In order to apply the same barriers in OpenMP code, we use \texttt{omp parallel for} by introducing a for loop code segment as shown between lines 5 and 8 in Listing~\ref{code:synchthread_ompt}.

To implement device-function-level synchronization in OpenMP, we decomposed target functions into sets of subroutines to enable thread synchronization, and call these subroutines in a global function.

% \lstset{language=C++,basicstyle=\footnotesize,numberstyle=\tiny,frame=single,caption={A CUDA device function which requires thread synchronization.},label=code:device_fn_cuda}
% \lstinputlisting{listings/devicefn_cuda.cpp}

% Listing~\ref{code:device_fn_cuda} shows a simplified example of a device function in CUDA, to compare to a corresponding OpenMP code segment in a global function shown in Listing~\ref{code:target_fn_ompt}, with a target teams construct. %The \texttt{device\_function} requires two thread barriers. 

%\lstset{language=C++,basicstyle=\footnotesize,numberstyle=\tiny,frame=single,caption={A global function in OpenMP target offloading.},label=code:global_fn_ompt}
%\lstinputlisting{listings/global_fn_ompt.cpp}\vspace{-1em}

\lstset{language=C++,basicstyle=\footnotesize,numberstyle=\tiny,frame=single,caption={A code segment that consist of decomposed target functions, based on the need for synchronization},label=code:target_fn_ompt}
\begin{lstlisting}
    float energy = 0.0f;
#pragma omp parallel for
    for (int atom_id = 0; atom_id < natoms; atom_id++) {
        get_atompos( atom_id, ... );
    }
    for(int rotcyc=0; rotcyc < nrotcyc; rotcyc++){
        int start = rot*work_pteam;
        int end = start +work_pteam;
        if ( end > rot_length ) end = rot_length; 
#pragma omp parallel for  
        for (int rot = start; rot  < end; rotr++){
            rotate_atoms(rot ...);
        }// end for rotations; a rotation cycle
    }// end for rotation cycles
    // team-level reduction
#pragma omp parallel for reduction(+:energy)
    for (int atom_id = 0; atom_id < natoms; atom_id++){
        energy += calc_erenergy( atom_id, ... );
    } 
\end{lstlisting}

% \lstset{language=C++,basicstyle=\footnotesize,numberstyle=\tiny,frame=single,caption={Thread synchronization in CUDA.},label=code:device_fn_cuda}
% \lstinputlisting{listings/device_fn_cuda.cpp}
%The device function is split into three subroutines (target functions \texttt{target\_fn\_1}, \texttt{target\_fn\_2}, and \texttt{target\_fn\_3} ) shown in Listing~\ref{code:target_fn_ompt}, each with \texttt{omp parallel for} regions each of which provide implicit barriers. In the miniMDock code, the device function \texttt{calc\_energy} was re-implemented in this manner for the OpenMP offload version.
The rotations of atoms are performed by completing rotation cycles in a particular order, therefore the CUDA code uses \texttt{\_\_synchthreads()}.
%as shown in Listing~\ref{code:device_fn_cuda} (in line 8). 
It treats the number of rotation cycles \texttt{nrotcyc} implicitly for the \texttt{rot\_length} ( = \texttt{nrotcyc} x \texttt{block size}) and makes sure that all threads complete their rotation before moving to the next cycle. In order to implement this rotation task in OpenMP, we use a nested for loop with \texttt{pragma omp parallel for} over the inner loop to create a team of threads for a cycle of rotation (lines between 6 and 14 in Listing~\ref{code:target_fn_ompt}).    

\subsection{Approach 2: Using ``teams" construct}

Another approach which is closer to a direct conversion from the CUDA version, is \textit{currently} more portable over the different compilers because it uses only the older and more fundamental features within OpenMP, but is less desirable in that it does not make use of the more human-readable, closer-to-the-algorithm high-level style that compiler directives provide, but requires a more hardware-level programming style. We used the \texttt{omp parallel} construct to manually specify shared and distributed work over threads similar to the CUDA version. 

Listing~\ref{code:synchthread_omp_par} shows the equivalent OpenMP code, written in this approach, for the CUDA kernel shown in Listing~\ref{code:synchthread_cuda}. One benefit of this method is direct-handling of thread synchronization; we can use \texttt{pragma omp barrier} for thread synchronization directly. 
%even for the device functions, which was not possible in the previous approach.
 \lstset{language=C++,basicstyle=\footnotesize,numberstyle=\tiny,frame=single,caption={Thread synchronization in OpenMP target offloading using \textit{Approach 2}.},label=code:synchthread_omp_par}
\begin{lstlisting}
void gpu_kernel( uint32_t nteams, uint32_t threads_pteam, parameters ... ){
#pragma omp target teams num_teams(nteams) thread_limit(threads_pteam)
    {
		float3_struct A[N];  // shared data
	#pragma omp parallel
	 {
	     const int threadIdx = omp_get_thread_num();
	     const int blockDim = omp_get_num_threads();
	     const int blockIdx = omp_get_team_num();
	     const int gridDim = omp_get_num_teams();
	     for (uint32_t idx = blockIdx; idx < nteams; idx+=gridDim) {  // for teams
	        float  x = device_function( A, ... );
            #pragma omp barrier
            compute( ... );
            f (threadIdx.x == 0) { 
        	    //Work for the master thread
        	    }
	        }
	    } // end of parallel region
	} // end of teams region
}
\end{lstlisting}
%\subsection{Deep copying}
%Copying complicated data structures to the device can require a significant amount of code restructuring as well. Listing~\ref{code:data_struct} shows a simplified example of a type definition for a structure of static arrays, \texttt{struct\_static\_array} and how it is used it as a pointer in another structure, \texttt{Data}. 

%\lstset{language=C++,basicstyle=\footnotesize,numberstyle=\tiny,frame=single,caption={A complicated data structure, using a struct of static array as a pointer in a struct.},label=code:data_struct}
%\lstinputlisting{listings/data_struct.cpp}

%Listing~\ref{code:deepcpy_ompt} shows that how these kind of data structures are mapped to the device with OpenMP and how updates from the host are performed with directives. 

%\lstset{language=C++,basicstyle=\footnotesize,numberstyle=\tiny,frame=single,caption={Mapping and deep copying struct of static array to device using OpenMP target offloading.},label=code:deepcpy_ompt}
%\lstinputlisting{listings/deepcpy_ompt.cpp}

Even with the choice to maintain as much code structure as possible from the original low-level architecture-specific code such as CUDA (which may potentially not result in the most optimal performance), there is a significant amount of work required to rewrite the program into a higher-level version appropriate for OpenMP offloading, including the careful restructuring of thread barriers and synchronization, of device functions, and of the deep copies of complex structures onto the device. This effort may be greater than the effort required for adding directive-based offload to an existing program written for the CPU. However, with the hope that the resulting version will be portable to multiple architectures, this time investment may be worthwhile if the performance of the portable version is close to that of the architecture-specific version. In the next section, we present performance results across compilers and GPU devices, both AMD and NVIDIA GPUs.

% \subsection{Time for porting and productivity considerations}
\section{Experimental Performance Testing}
\subsection{Summary of compilers and systems tested}
We ran tests over four compilers and two GPU architectures (AMD and NVIDIA), using NVHPC for CUDA on NVIDIA, ROCm for HIP on AMD, and for OpenMP target offload, two compilers for NVIDIA GPUs: LLVM and NVHPC and three compilers for AMD GPUs: ROCm, OpenMP focused LLVM-Clang based AOMP, and HPE-Cray's Cray Compiling Environment (CCE). Values presented are means over 10 replica runs for each test. For NVIDIA testing, we used the Summit supercomputer, and for AMD testing, we used the Spock system. Both systems are housed at the Oak Ridge Leadership Computing Facility (OLCF). Summit is an IBM system containing approximately 4,600 IBM Power System AC922 compute nodes. Each node contains two IBM POWER9 processors and 6 NVIDIA Tesla V100 accelerators. Each processor is connected via dual NVLINK connections capable of a 25GB/s transfer rate in each direction.
\footnote{\url{https://docs.olcf.ornl.gov/systems/summit_user_guide.html}} 
Spock is an early-access testbed for the upcoming OLCF Frontier exascale supercomputer, which contains 36 compute nodes each with a 64-core AMD EPYC 7662 CPU and four AMD MI100 GPUs. The CPU is connected to all GPUs via PCIe Gen4, with 32 GB/s transfer rate in each direction.
\footnote{\url{https://docs.olcf.ornl.gov/systems/spock_quick_start_guide.html}}. On Summit we used the NVHPC 21.11 compiler as well as the LLVM 14.0 and 15.0 main branch development snapshots.\footnote{(git {\tt 7bdce6bcbda3a8b7dcdac6a7d8fb1083912830d7)}} On Spock we used the rocm 4.5, aomp/14.0.1\footnote{\url{https://github.com/RadeonOpenCompute/llvm-project/tree/aomp-14.0-1}}, and the CCE/12.0.1 compiler suites.

% \begin{table}[htbp]
% \caption{Compilers and systems tested}
% \begin{center}
% \begin{tabular}{ccc}
% \hline
% \textbf{Device}&\textbf{NVIDIA}&\textbf{AMD} \\
% \hline
% \textbf{Compiler}& nvhpc&ROCm \\
% \textbf{API}& CUDA&HIP\\
% \hline
% \textbf{Compiler}& clang&CCE \\
% \textbf{API}& CUDA&HIP\\
% \hline
% \textbf{Compiler}& nvhpc&ROCm \\
% \textbf{API}& OpenMP&OpenMP\\
% \hline
% \textbf{Compiler}& clang&CCE \\
% \textbf{API}& OpenMP&OpenMP\\
% \hline
% \end{tabular}
% \label{tab:systems}
% \end{center}
% \vspace{-1em}
% \end{table}
\subsection{Results}
 The two OpenMP programming approaches introduced above will be denoted as labeled omp\_dist, for approach 1 (\texttt{distribute} construct-based automatic work sharing) and omp\_par, for approach 2, (\texttt{parallel} construct-based manually specified work sharing).
 
\begin{figure}[htbp]
\centering{\includegraphics[ width=\linewidth]{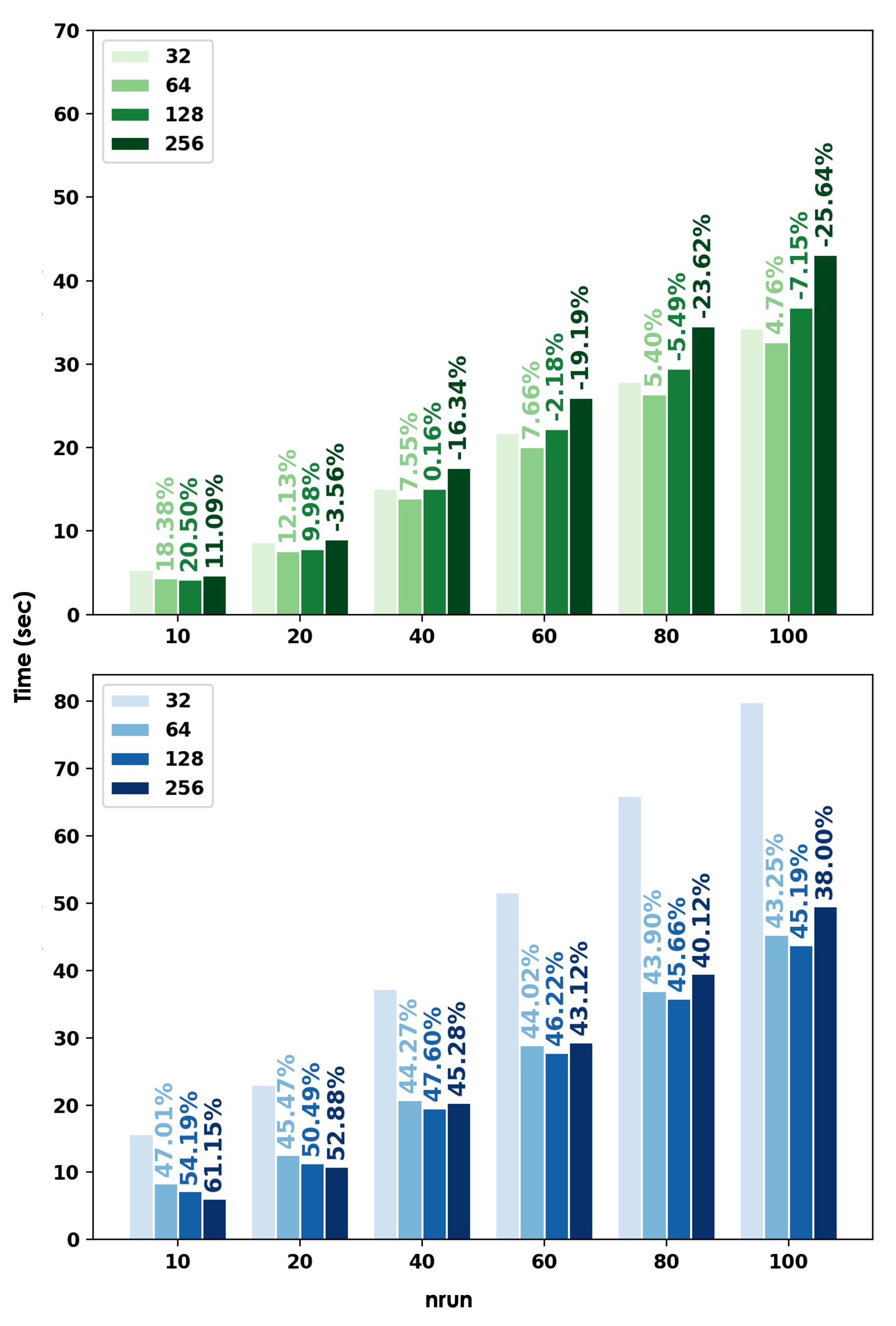}}
\caption{Effect of tuning threads-per-block parameter in CUDA on the NVIDIA V100 GPU (top panel) and threads-per-workgroup parameter in HIP on the AMD MI100 (bottom panel) for the large input. Shown is runtime in seconds for variations of this parameter and \texttt{nruns}, using values of the threads per block parameter of 32, 64, 128 and 256. }
\label{fig:cuda_hip_tpb}
\end{figure}

In order to establish the baseline with which to determine the performance portability of the target offload version, a best-performing value must be established for the non-portable code. There may be several parameters that can be tuned to optimize code performance. For both the CUDA and HIP versions using V100 and MI100 GPUs, respectively, we noted that tuning a variable within the code designating the number of threads-per-block (in CUDA) or threads-per-workgroup (in HIP) affected the performance of the program, with the most significant effects being on performance with the large input. Figure~\ref{fig:cuda_hip_tpb} displays these variations for CUDA on the NVIDIA V100 GPU and for HIP on the AMD MI100. The CUDA version is less sensitive to this parameter, and while the optimal value of threads-per-block provides a non-trivial improvement (up to 1.3$\times$ speedup) for CUDA, for HIP, performance is much more sensitive to to the analogous threads-per-workgroup value, with larger values providing 2-3$\times$ speedup. Also notable is that for HIP, larger speedups from increasing this value above 32 were obtained for smaller loads (close to 3$\times$ for \texttt{nruns}=10 compared to less than 2$\times$ for \texttt{nruns}=100). 

\begin{figure}[htb]
\centering{\includegraphics[ width=\linewidth]{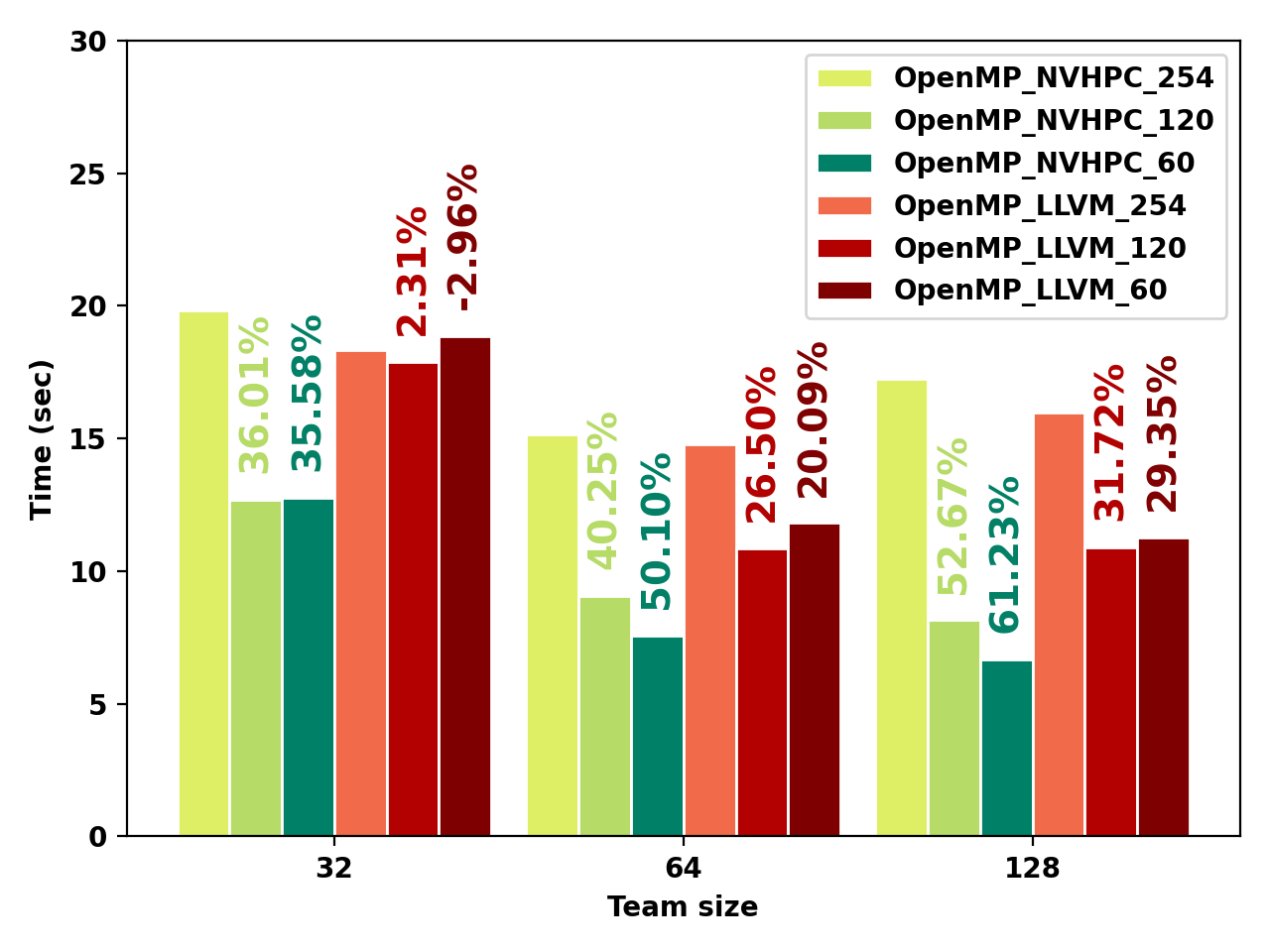}}
\caption{Effects of tuning \texttt{thread\_limit} and \texttt{maximum register count} on run time for the large input for OpenMP using NVHPC and LLVM compilers on NVIDIA V100 GPUs.}
\label{fig:nvhpc_llvm_tuning}
\vspace{-1em}
\end{figure}

Compiler-level optimizations are also possible for OpenMP GPU offload, and should be considered while assessing performance portability of a directive-based solution. Figure~\ref{fig:nvhpc_llvm_tuning} shows the effects of the tuning the \texttt{thread\_limit} and \texttt{maximum register count per thread} parameters on performance with OpenMP-NVHPC using the omp\_par control flow. Tuning the maximum register count from the default value, 254, to 60 shows significant performance improvements for each thread\_limit; the effect grows from smaller thread\_limit 32 at 35.58\%, to 61.23\% relative speedup for 128. The thread\_limit parameter has a similar effect as the threads-per-block/workgroup parameter in the CUDA/HIP versions. The increase of \texttt{thread\_limit} value to 128 results in a significant improvement in performance, about 91.29\% relative speedup compared to using the original setting of 32 for the best choice of maximum register count 60. Tuning both of these parameters brings the comparative slowdown with respect to the CUDA version on the V100 GPU down to 1.52$\times$ from the almost 4.51$\times$ for the large input. LLVM-clang gives the best performance for maximum register count 60 and \texttt{thread\_limit} 64. Tuning both parameters bring the comparative slowdown with respect to the CUDA version down to 2.47$\times$ from the almost 4.16$\times$ for the large input. 

\begin{figure}[htb]
\centering{\includegraphics[ width=0.9\linewidth]{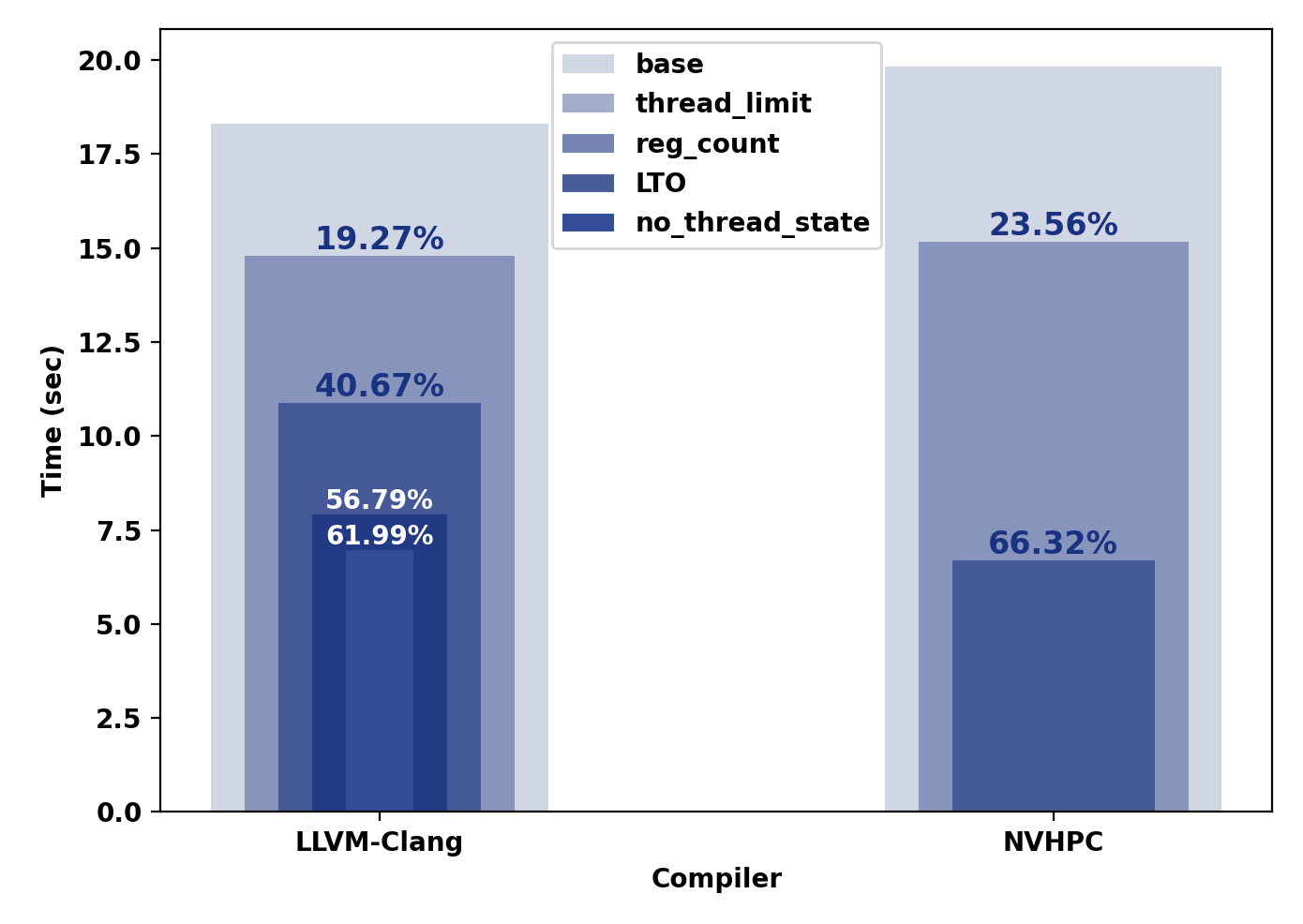}}
\caption{Effects of the tuning parameters on the performance of the program for the large input for OpenMP on NVIDIA V100 GPUs using LLVM and NVHPC compiler. Shown is runtime in seconds for variations of \texttt{thread\_limit}. }
\label{fig:omp_tune}
\vspace{-1em}
\end{figure}

\begin{figure}[htb]
\centering{\includegraphics[ width=0.9\linewidth]{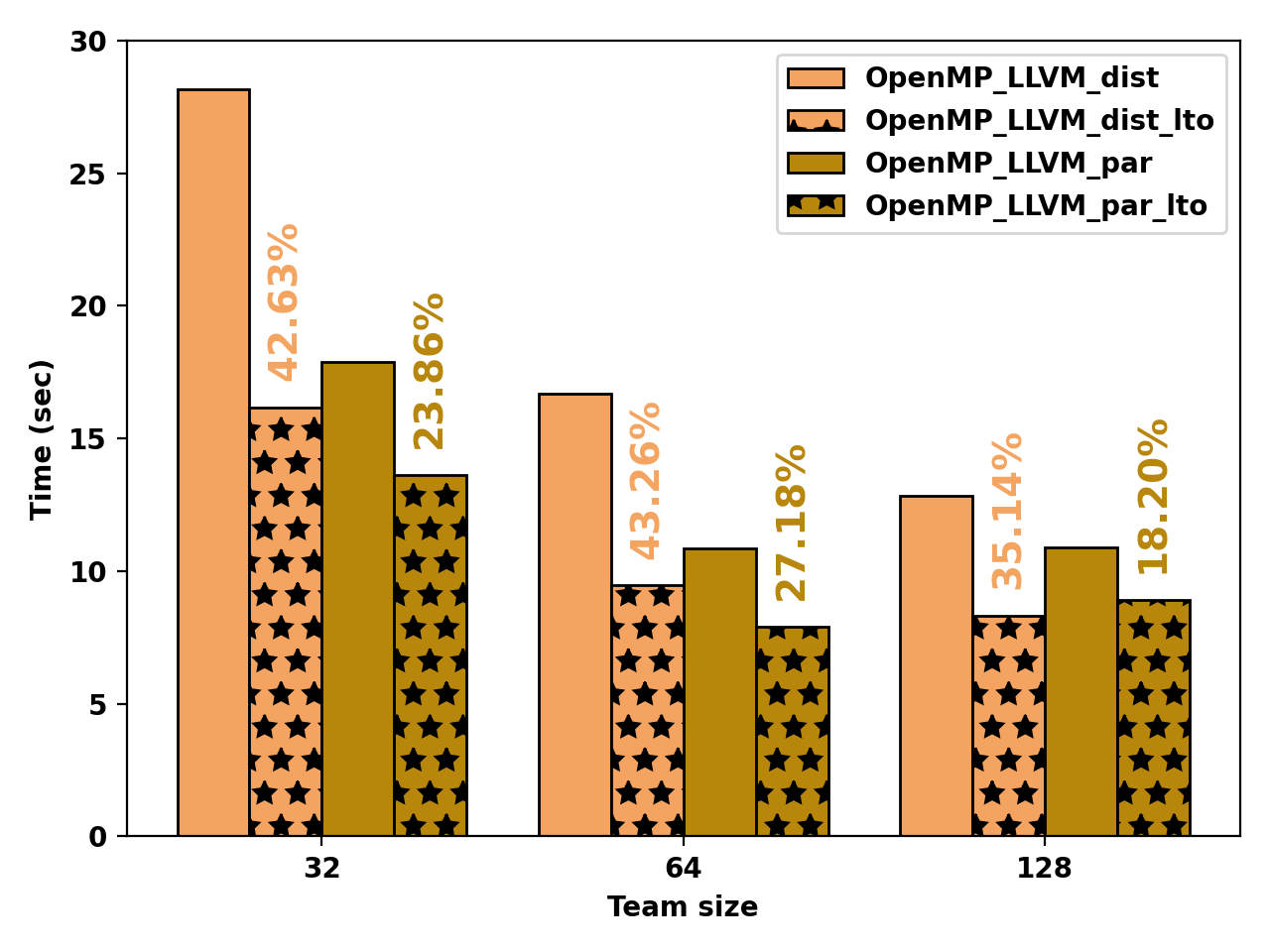}}
\caption{Effects of LLVM's \texttt{link time optimization} feature for the large input for OpenMP using LLVM compiler on NVIDIA V100 GPUs, for both program strategies. Shown is runtime in seconds for variations of \texttt{thread\_limit}. }
\label{fig:llvm_lto}
\vspace{-1em}
\end{figure}

An additional optimizations, recently  added to LLVM, are the use of link-time optimization (LTO),\footnote{https://github.com/llvm/llvm-project/commit/  2f9ace9e9a5816684b3c19528bd4a3908b2b8ac0} and no thread state which was used with the latest LLVM version 15. Figure~\ref{fig:omp_tune} shows the cumulative performance improvements for LLVM-Clang and NVHPC compilers for the various compiler optimizations, including the use of LTO for LLVM, with the omp\_par control flow. The LTO optimization helps bring LLVM closer to the optimized performance that NVHPC achieves. Additional LLVM optimizations for OpenMP offload on NVIDIA GPUs are currently in development. The effects of this optimization on the V100 are also shown in Figure~\ref{fig:llvm_lto}, showing significant performance increase for both strategies, around 43.26\% for omp\_dist and 27.18\% for omp\_par, using thread\_limit 64. LTO reduces slowdown from 2.92$\times$ to 1.89$\times$ and 2.47$\times$ to 1.80$\times$ for the two versions, respectively, with respect to the CUDA version. %Despite this additional optimization, LLVM still provides reduced performance compared to NVHPC, however the active development of this implementation suggest that the gap may soon be closed.

\begin{figure}[htbp]
\centering{\includegraphics[ width=0.9\linewidth]{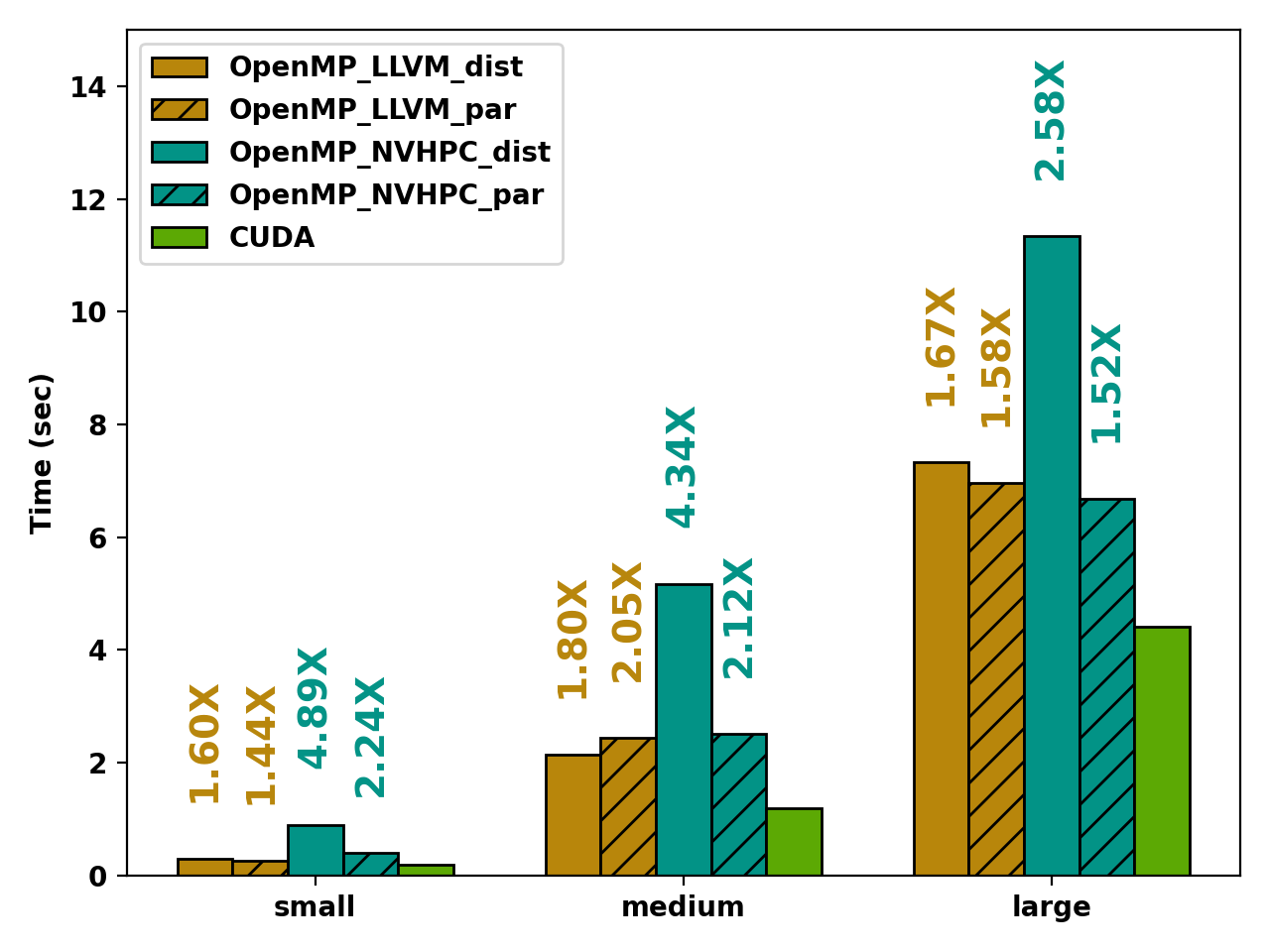}}
\caption{Performance of OpenMP target offload on NVIDIA V100 GPUs using NVHPC and LLVM compilers for the two different programmatic strategies, teams distribute (dist, approach 1) and teams with parallel for (par, approach 2). Indicated are times in seconds, and slowdown ($\times$) compared to the CUDA version for each input (small, medium, and large ligand);\texttt{nruns}=10. }
\label{fig:cuda_omp_10}
\vspace{-1em}
\end{figure}

Figure~\ref{fig:cuda_omp_10} shows the performance of the two OpenMP programmatic approaches, omp\_dist (approach 1), and omp\_par (approach 2), using LLVM-clang OpenMP target offload and the NVIDIA HPC SDK NVHPC OpenMP target offload, compared with the CUDA version, for \texttt{nruns}=10. Here we used the optimal combination of compiler tuning strategies described above, in all cases. The CUDA-style manually-specified work sharing strategy (omp\_par, approach 2) consumes less time than the automatic version (omp\_dist, approach 1) for both LLVM-clang and NVHPC compilers, but with a much smaller gap for LLVM. As mentioned above, reducing compiler sensitivity to program control flow with respect to performance is a desired approach to enable high-level programming APIs to maximize productivity.

For the LLVM omp\_dist version, the maximum slowdown compared to CUDA is for the small input, at 1.6$\times$ slower, while for the medium and large inputs this declines to 1.8$\times$ and 1.67$\times$. The NVHPC omp\_dist version follows the same pattern: the slowdown gets progressively better as the input size increases, from 4.89$\times$ for the small input, to 2.58$\times$ for the large one. A similar pattern is found for NVHPC with omp\_par for NVHPC, but not for LLVM: for the small input it is 1.44$\times$ slower, while for the medium and large inputs this grows to 2.05$\times$ and 1.58$\times$. These results indicate that both implementations are sensitive to the program control flow, but that LLVM is less so. 

%\begin{figure}[htb]
%\centering{\includegraphics[ width=\linewidth]{img/omp_llvm_tune.png}}
%\caption{Tuning performance of the program with \texttt{thread\_limit} and \texttt{maximum register count} for the large input for OpenMP using LLVM compiler on Summit. Shown is runtime in seconds for variations of \texttt{thread\_limit} and \texttt{maximum register count}. }
%\label{fig:llvm_tuning}
%\vspace{-1em}
%\end{figure}

In either case, the performance can be considered significantly less than the CUDA version, indicating that true performance portability has not yet been achieved for NVIDIA GPUs despite trying different code structuring and compiler-level optimizations. Our initial speculation was that due to the code structure being originally designed for CUDA, these codes patterns were sub-optimal for OpenMP offloading, and that loop structures and data patterns would have to be rearranged. However, results on AMD GPUs seemed to contradict this conclusion.

\begin{figure}[htbp]
\centering{\includegraphics[ width=\linewidth]{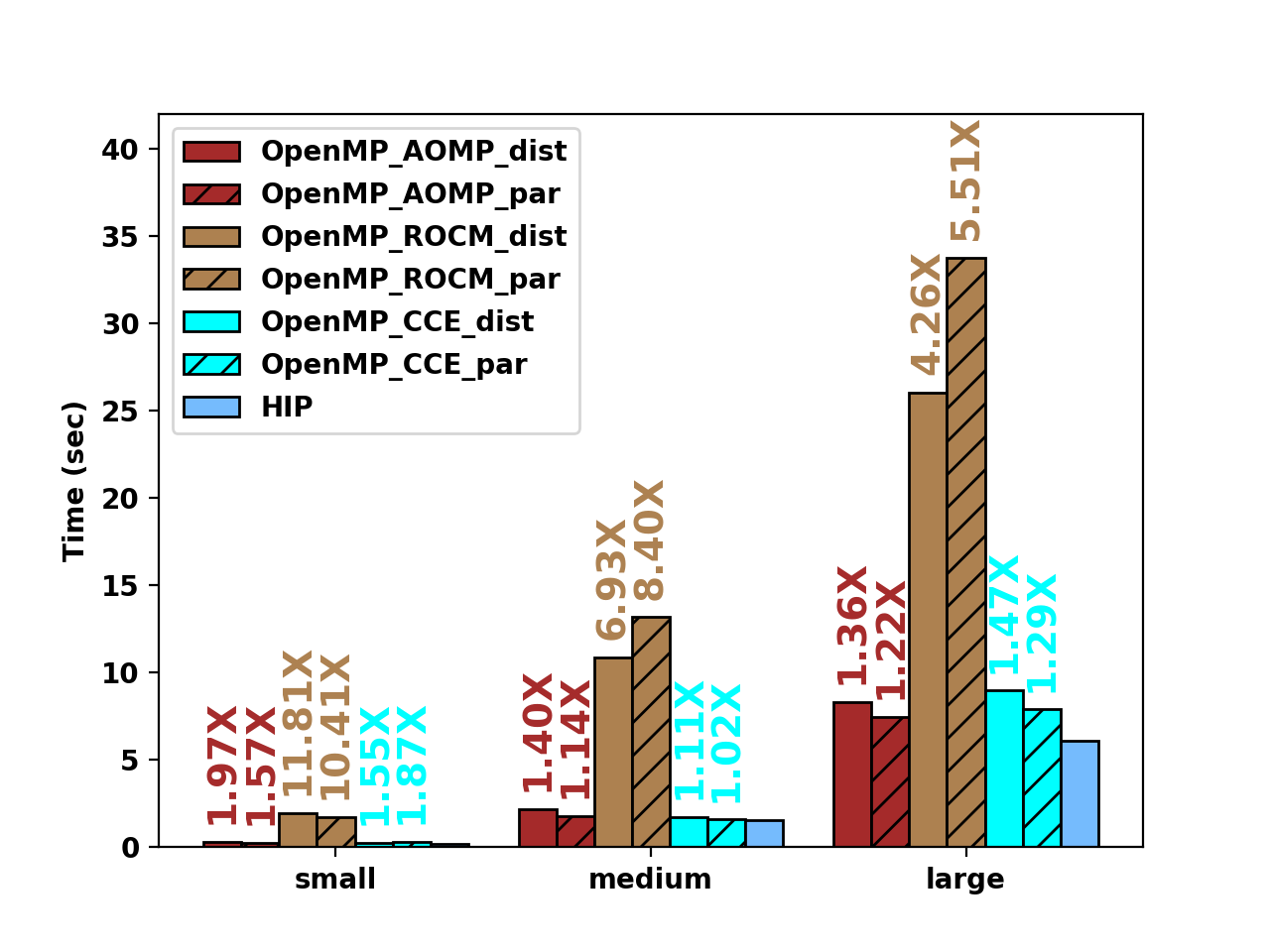}}
\caption{Performance of OpenMP target offload on AMD MI100 GPUs for \texttt{nruns} set to 10. Indicated are the exact times in seconds, and the slowdown ($\times$) compared to the HIP version for each input (small, medium, and large ligand). }
\label{fig:hip_omp_10}
\vspace{-1em}
\end{figure}

Figure~\ref{fig:hip_omp_10} shows performance on AMD MI100 GPUs, using HIP and the OpenMP offload versions of the miniapp with the HPE-Cray CCE compiler and with AMD's ROCm and AOMP compilers. Notable is a narrower gap in performance between HIP and the OpenMP version with the CCE compiler for the medium and large inputs. For the medium input, performance is nearly identical, which is an outstanding result. For the small and large input, the gap increases, resulting in a 1.55$\times$ and 1.47$\times$ slowdown, respectively, for omp\_dist using CCE and 1.87$\times$ and 1.29$\times$ slowdown, respectively, for omp\_par using CCE. For the best version of the large and medium input, this is within an acceptable range for a portable solution, while performance with the small input could be improved. The tuning of threads-per-workgroup with the CCE compiler did not result in any improvements (in fact, performance was slightly decreased) and gives the best performance at the default threads-per-workgroup value of 64. The AOMP compiler for the OpenMP version shows slightly better performance than CCE for the large input, and gives the minimum slowdown for that input, 1.22$\times$. Performance of the ROCm compiler for the OpenMP version was considerably worse, and in contrast to all other compilers it shows better performance for omp\_dist strategy than omp\_par strategy in some cases. However, the performance of the AOMP compiler shows advances that will soon be incorporated into the ROCm compiler, and thus provides a more favorable view of the future ROCm stack.
\begin{figure}[htbp]
\centering{\includegraphics[ width=\linewidth]{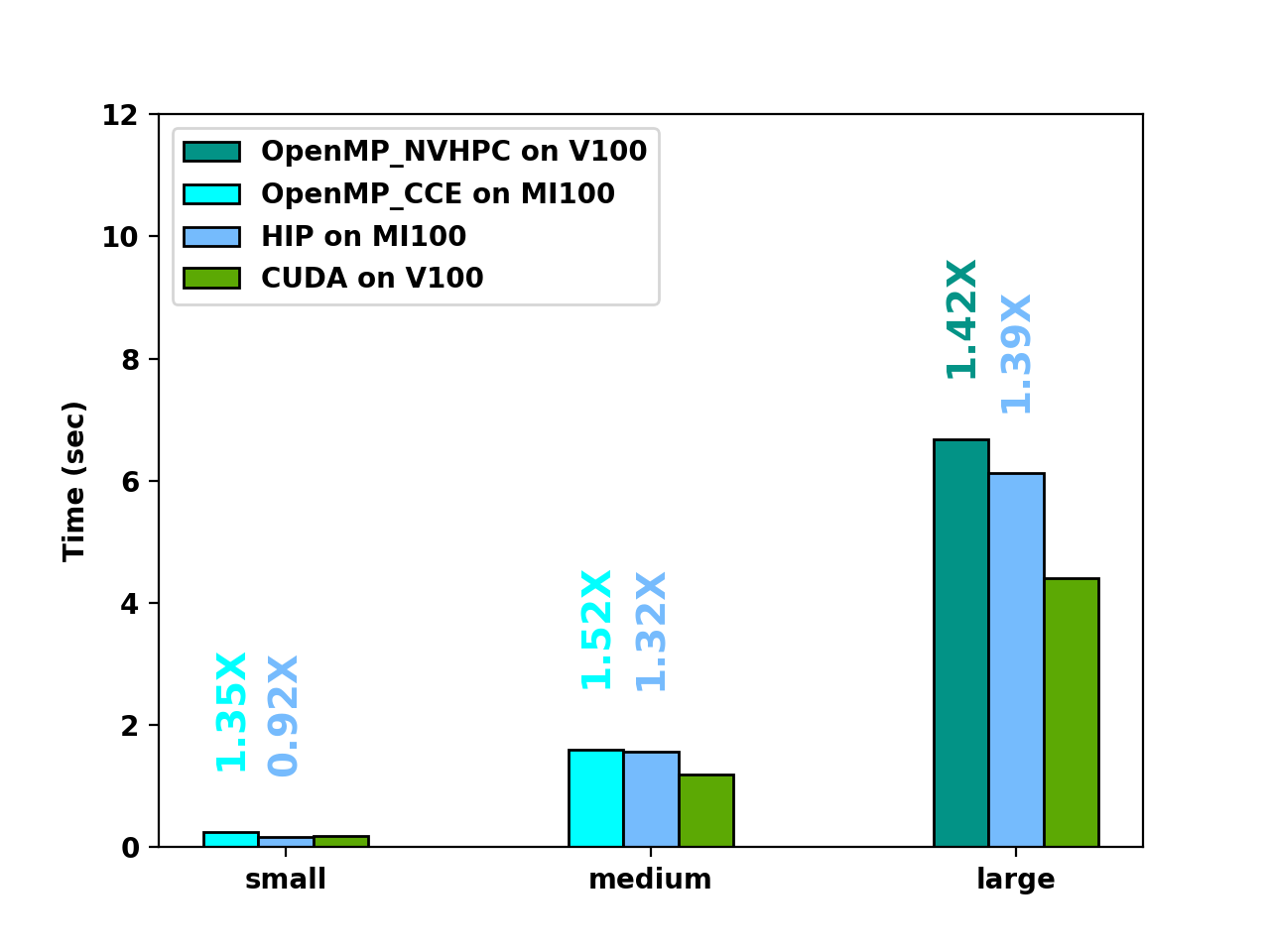}}
\caption{Performance of the three best combinations of API, compiler and device across both systems with \texttt{nruns} set to 10. CUDA: CUDA on NVIDIA V100 GPU with NVHPC compiler. HIP: HIP on AMD MI100 GPU with ROCm compiler. OpenMP-CCE: OpenMP target offload on AMD MI100 GPU with CCE compiler. OpenMP-NVHPC: OpenMP target offload on NVIDIA V100 GPU with NVHPC compiler. Indicated is the performance ratio ($\times$) compared to CUDA for each input (small, medium, and large ligand). }
\label{fig:best3}
\end{figure}
 Figure~\ref{fig:best3} shows a summary of the best performing versions across the two GPUs (NVIDIA and AMD), with the CUDA, HIP, and OpenMP target offloading results for \texttt{nruns} set to 10 across the input sizes. We see that for the small input size, the HIP version actually gets better performance than CUDA. In the other cases, CUDA is best. OpenMP-LLVM on the NVIDIA GPU and OpenMP-CCE on the AMD GPU provide identical performance for the small input, but for the medium input size, OpenMP-CCE provides better performance than OpenMP-LLVM, and shows a similar to HIP with respect to CUDA. For HIP, there is a similar relative slowdown for the medium and large inputs. OpenMP-NVHPC and OpenMP-AOMP give better performance among the OpenMP compilers for the large input, with OpenMP-NVHPC on NVIDIA GPU outperforming OpenMP-AOMP on AMD.

We therefore see that a compiler-directive-based solution for a performance portable molecular docking program is becoming more of a viable reality. This implies the possibility of maintaining a single version of a program that can run on multiple GPU architectures, in a rapidly changing landscape for cloud servers and HPC clusters where several new vendors are entering the GPU arena. Our results also indicate that in general, compiler-based GPU offloading solutions for other types of programs may prove to be practical. However, it is clear that careful tuning of available compiler optimization parameters, choice of compiler, and even rearrangements in programmatic control flow may be necessary to achieve optimal performance with the directive approach. In principle, such considerations detract from productivity, which is one aim of performance portability efforts. Luckily, tools for auto-tuning performance are a focus of HPC research, and may provide assistance in reducing time spent tuning for compiler-based solutions.

\section{Discussion and Conclusion}
Here we tested the miniMDock molecular docking miniapp on NVIDIA and AMD GPUs using both native APIs for each device and new compiler-directive strategies. As a basline, performance of the HIP version on AMD GPUs was slightly faster than the CUDA version on NVIDIA GPUs for our small input, and under 1.4$\times$ slower for the other two inputs, after tuning of thread-grouping parameters. We found that for miniMDock, it is possible to port an architecture-specific GPU-accelerated program to a more portable OpenMP offload version that can run on two different GPU architectures while avoiding a complete re-writing of the program, and achieve comparable performance to the original version-- depending on the underlying decisions made by the compiler. Care must be taken to correctly translate thread synchronization and the placement of implicit barriers. The resulting program more closely reflects the scientific problem and less the architecture of the device, and can make use of multiple levels of nested parallelism explicitly, via constructs provided by OpenMP. We found that for the available compilers that support GPU offloading using OpenMP, truly comparable performance is not yet achieved for NVIDIA GPUs, with slowdowns compared to the CUDA version exceeding 1.5$\times$ for all input sizes tested, and despite tuning of several compiler optimization parameters and testing two different programmatic patterns. Performance of OpenMP offloading using the HPE-Cray CCE compiler resulted in a slowdown of 1.3$\times$ for the large input compared to HIP, and almost identical performance for the medium input, compared to the HIP version. This indicates that true performance portability using directives is becoming increasingly more realizable. It should be noted, however, that controls such as the thread limit and maximum register count do have a major impact on performance of higher-level programming models, as does the choice of compiler. Our results thus demonstrate the importance of the compiler back-ends for enabling truly performance-portable programs with directive-based offloading across different GPU vendors, and provide an optimistic outlook for performance-portable solutions as new GPU architectures enter the HPC and cloud ecosystems.

\section{Acknowledgment}
This research used resources of the Oak Ridge Leadership Computing Facility, which is a DOE Office of Science User Facility supported under Contract DE-AC05-00OR22725. We thank Oscar Hernandez for valuable discussions.

\bibliographystyle{plain}
\bibliography{miniMDock}

\begin{thebibliography}{10}

\bibitem{adjoua2021tinker}
Olivier Adjoua, Louis Lagard{\`e}re, Luc-Henri Jolly, Arnaud Durocher, Thibaut
  Very, Isabelle Dupays, Zhi Wang, Th{\'e}o~Jaffrelot Inizan, Fr{\'e}d{\'e}ric
  C{\'e}lerse, Pengyu Ren, et~al.
\newblock Tinker-hp: Accelerating molecular dynamics simulations of large
  complex systems with advanced point dipole polarizable force fields using
  gpus and multi-gpu systems.
\newblock {\em Journal of chemical theory and computation}, 17(4):2034--2053,
  2021.

\bibitem{boehm2018evaluating}
Swen Boehm, Swaroop Pophale, Ver{\'o}nica G~Vergara Larrea, and Oscar
  Hernandez.
\newblock Evaluating performance portability of accelerator programming models
  using spec accel 1.2 benchmarks.
\newblock In {\em International Conference on High Performance Computing},
  pages 711--723. Springer, 2018.

\bibitem{daley2020case}
Christopher Daley, Hadia Ahmed, Samuel Williams, and Nicholas Wright.
\newblock A case study of porting hpgmg from cuda to openmp target offload.
\newblock In {\em International Workshop on OpenMP}, pages 37--51. Springer,
  2020.

\bibitem{drews2000drug}
Jürgen Drews.
\newblock Drug discovery: A historical perspective.
\newblock {\em Science}, 287(5460):1960--1964, 2000.

\bibitem{gayatri2018case}
Rahulkumar Gayatri, Charlene Yang, Thorsten Kurth, and Jack Deslippe.
\newblock A case study for performance portability using openmp 4.5.
\newblock In {\em International Workshop on Accelerator Programming Using
  Directives}, pages 75--95. Springer, 2018.

\bibitem{glaser2021high}
Jens Glaser, Josh~V Vermaas, David~M Rogers, Jeff Larkin, Scott LeGrand, Swen
  Boehm, Matthew~B Baker, Aaron Scheinberg, Andreas~F Tillack, Mathialakan
  Thavappiragasam, et~al.
\newblock High-throughput virtual laboratory for drug discovery using massive
  datasets.
\newblock {\em The International Journal of High Performance Computing
  Applications}, page 10943420211001565, 2021.

\bibitem{gorgulla2020open}
Christoph Gorgulla, Andras Boeszoermenyi, Zi-Fu Wang, Patrick~D Fischer, Paul~W
  Coote, Krishna M~Padmanabha Das, Yehor~S Malets, Dmytro~S Radchenko, Yurii~S
  Moroz, David~A Scott, et~al.
\newblock An open-source drug discovery platform enables ultra-large virtual
  screens.
\newblock {\em Nature}, 580(7805):663--668, 2020.

\bibitem{harrell2018effective}
Stephen~Lien Harrell, Joy Kitson, Robert Bird, Simon~John Pennycook, Jason
  Sewall, Douglas Jacobsen, David~Neill Asanza, Abaigail Hsu, Hector~Carrillo
  Carrillo, Hessoo Kim, et~al.
\newblock Effective performance portability.
\newblock In {\em 2018 IEEE/ACM International Workshop on Performance,
  Portability and Productivity in HPC (P3HPC)}, pages 24--36. IEEE, 2018.

\bibitem{hsu2018performance}
Abigail Hsu, David~Neill Asanza, Joseph~A Schoonover, Zach Jibben, Neil~N
  Carlson, and Robert Robey.
\newblock Performance portability challenges for fortran applications.
\newblock In {\em 2018 IEEE/ACM International Workshop on Performance,
  Portability and Productivity in HPC (P3HPC)}, pages 47--58. IEEE, 2018.

\bibitem{juckeland2016describing}
Guido Juckeland, Oscar Hernandez, Arpith~C Jacob, Daniel Neilson, Ver{\'o}nica
  G~Vergara Larrea, Sandra Wienke, Alexander Bobyr, William~C Brantley, Sunita
  Chandrasekaran, Mathew Colgrove, et~al.
\newblock From describing to prescribing parallelism: Translating the spec
  accel openacc suite to openmp target directives.
\newblock In {\em International Conference on High Performance Computing},
  pages 470--488. Springer, 2016.

\bibitem{legrand2020gpu}
Scott LeGrand, Aaron Scheinberg, Andreas~F Tillack, Mathialakan
  Thavappiragasam, Josh~V Vermaas, Rupesh Agarwal, Jeff Larkin, Duncan Poole,
  Diogo Santos-Martins, Leonardo Solis-Vasquez, Andreas Koch, Stefano Forli,
  Oscar Hernandez, Jeremy~C Smith, and Ada Sedova.
\newblock {GPU-Accelerated Drug Discovery with Docking on the Summit
  Supercomputer: Porting, Optimization, and Application to COVID-19 Research}.
\newblock In {\em Proc. 11th ACM Int. Conf. Bioinformatics, Comput. Biol. Heal.
  Informatics}, 2020.

\bibitem{lopez2016towards}
M~Graham Lopez, Ver{\'o}nica~Vergara Larrea, Wayne Joubert, Oscar Hernandez,
  Azzam Haidar, Stanimire Tomov, and Jack Dongarra.
\newblock Towards achieving performance portability using directives for
  accelerators.
\newblock In {\em 2016 Third Workshop on Accelerator Programming Using
  Directives (WACCPD)}, pages 13--24. IEEE, 2016.

\bibitem{martineau2016pragmatic}
Matt Martineau, James Price, Simon McIntosh-Smith, and Wayne Gaudin.
\newblock Pragmatic performance portability with openmp 4. x.
\newblock In {\em International Workshop on OpenMP}, pages 253--267. Springer,
  2016.

\bibitem{messer2018miniapps}
OE~Bronson Messer, Ed~D’Azevedo, Judy Hill, Wayne Joubert, Mark Berrill, and
  Christopher Zimmer.
\newblock {MiniApps} derived from production {HPC} applications using multiple
  programming models.
\newblock {\em The International Journal of High Performance Computing
  Applications}, 32(4):582--593, 2018.

\bibitem{Openmp5}
OpenMP.
\newblock {OpenMP 5.0 Reference Guide}.
\newblock
  \url{https://www.openmp.org/wp-content/uploads/OpenMPRef-5.0-1119-01-TSK-web.pdf}.

\bibitem{pagadala2017software}
Nataraj~S Pagadala, Khajamohiddin Syed, and Jack Tuszynski.
\newblock Software for molecular docking: a review.
\newblock {\em Biophysical reviews}, 9(2):91--102, 2017.

\bibitem{pennycook2019implications}
Simon~J Pennycook, Jason~D Sewall, and Victor~W Lee.
\newblock Implications of a metric for performance portability.
\newblock {\em Future Generation Computer Systems}, 92:947--958, 2019.

\bibitem{santos2019d3r}
Diogo Santos-Martins, Jerome Eberhardt, Giulia Bianco, Leonardo Solis-Vasquez,
  Francesca~Alessandra Ambrosio, Andreas Koch, and Stefano Forli.
\newblock {D3R} grand challenge 4: prospective pose prediction of {BACE1}
  ligands with {AutoDock-GPU}.
\newblock {\em Journal of Computer-Aided Molecular Design}, 33(12):1071--1081,
  2019.

\bibitem{santos2019accelerating}
Diogo Santos-Martins, Leonardo Solis-Vasquez, Andreas Koch, and Stefano Forli.
\newblock Accelerating {AutoDock4} with {GPUs} and gradient-based local search.
\newblock {\em ChemRxiv}, 2019.

\bibitem{sedova2018high}
Ada Sedova, John~D Eblen, Reuben Budiardja, Arnold Tharrington, and Jeremy~C
  Smith.
\newblock High-performance molecular dynamics simulation for biological and
  materials sciences: challenges of performance portability.
\newblock In {\em 2018 IEEE/ACM International Workshop on Performance,
  Portability and Productivity in HPC (P3HPC)}, pages 1--13. IEEE, 2018.

\bibitem{sedova2018using}
Ada Sedova, Andreas~F Tillack, and Arnold Tharrington.
\newblock Using compiler directives for performance portability in scientific
  computing: kernels from molecular simulation.
\newblock In {\em International Workshop on Accelerator Programming Using
  Directives}, pages 22--47. Springer, 2018.

\bibitem{solis1981minimization}
Francisco~J Solis and Roger J-B Wets.
\newblock Minimization by random search techniques.
\newblock {\em Mathematics of operations research}, 6(1):19--30, 1981.

\bibitem{stone2010opencl}
John~E. Stone, David Gohara, and Guochun Shi.
\newblock Opencl: A parallel programming standard for heterogeneous computing
  systems.
\newblock {\em Computing in Science Engineering}, 12(3):66--73, 2010.

\bibitem{thavappir2021addressing}
Mathialakan Thavappiragasam, Vivek Kale, Oscar Hernandez, and Ada Sedova.
\newblock Addressing load imbalance in bioinformatics and biomedical
  applications: Efficient scheduling across multiple gpus.
\newblock In {\em 2021 IEEE International Conference on Bioinformatics and
  Biomedicine (BIBM)}, pages 1992--1999, 2021.

\bibitem{thavappiragasam2020performance}
Mathialakan Thavappiragasam, Aaron Scheinberg, Wael Elwasif, Oscar Hernandez,
  and Ada Sedova.
\newblock Performance portability of molecular docking miniapp on leadership
  computing platforms.
\newblock In {\em 2020 IEEE/ACM International Workshop on Performance,
  Portability and Productivity in HPC (P3HPC)}, pages 36--44. IEEE, 2020.

\bibitem{vermaas2020supercomputing}
Josh~Vincent Vermaas, Ada Sedova, Matthew~B Baker, Swen Boehm, David~M Rogers,
  Jeff Larkin, Jens Glaser, Micholas~D Smith, Oscar Hernandez, and Jeremy~C
  Smith.
\newblock Supercomputing pipelines search for therapeutics against covid-19.
\newblock {\em Computing in Science \& Engineering}, 23(1):7--16, 2020.

\end{thebibliography}
\end{document}